\documentclass[]{iopart}
\usepackage{iopams,graphicx,hyperref}
\usepackage[usenames,dvipsnames]{color}
\usepackage{mathrsfs}
\DeclareSymbolFontAlphabet{\mathrsfs}{rsfs}

\newcommand{\bea}{\begin{eqnarray}}
\newcommand{\eea}{\end{eqnarray}}
\newcommand{\be}{\begin{equation}}
\newcommand{\ee}{\end{equation}}
\newcommand{\no}{\nonumber}
\newcommand{\p}{\partial}

\begin{document}

\title{Hamiltonian structure and constraint algebra in the 
(2+2) formalism}
\author{J.H. Yoon}
\address{School of Physics,
Konkuk University, Seoul 143-701, Korea}
\ead{yoonjh@konkuk.ac.kr}
\date{\today}

\begin{abstract}
The canonical formalism of the (2+2) formulation of 
general relativity of 4 spacetime dimensions is studied under no symmetry assumptions, 
where the spacetime is viewed as a local product of a 2 dimensional base manifold of Lorentzian signature with the vertical space as its complement. 
The affine null parameter is chosen as the time coordinate whose level surfaces are
3 dimensional spacelike hypersurfaces.
From the first-order action principle, Hamilton's equations of 
motion and the constraints are obtained, which are found to be equivalent to the 
Einstein's equations. The constraint algebra is also presented,
which has interesting subalgebras such as the infinite dimensional Lie algebra of 
the diffeomorphisms of the 2 dimensional vertical space, infinite dimensional
Virasoro algebra associated with the 2 dimensional base manifold, 
and an analog of supertranslation. 
The symmetry algebra may be viewed as a generalization of the BMS or Spi group  
to a finite distance.

\end{abstract}


\section{Introduction}\label{s:intro}
Most researches of general relativity based on the null hypersurface formalism\cite{bondi62,sachs62,newman80} are based on 
the double null splitting of spacetime. The double null 
formalism\cite{newman62,GHP73,dinverno78,dinverno80,torre86,hayward93,hayward94} has considerable advantages in studying a number of classical problems associated with gravitational waves, whereas it has difficulties in constructing the canonical formalism and the initial value problem\cite{rendall90,choquet09}. On the other hand, the standard (3+1) formalism\cite{brill59,ADM60,york71,kuchar71,york72,choquet80,berger95} 
has a well-developed canonical formalism in which initial data are evolved along
a timelike direction, but it is not well-suited to physical situations where gravitational waves are important.

In this paper, I study the (2+2) bundle formalism\cite{yoon92,yoon93a,yoon99a,yoon04}
that incorporates advantages of both the null hypersurface formalism and the (3+1) formalism
under no symmetry assumptions. In this formalism, a spacetime of 4 dimensions
is viewed as a local product of a 2 dimensional base manifold of Lorentzian signature with the vertical space as its complement. 
General relativity in this (2+2) formalism appears a 2 dimensional 
gauge theory with the group of diffeomorphisms of the 2 dimensional vertical 
space as the gauge symmetry. From the perspective of the 4 dimensional spacetime,
the splitting of the spacetime into the 2 dimensional base manifold and the vertical space is quite arbitrary, so that the splitting of spacetime can be chosen in a such way 
that the resulting theory comprises the merits of both formalisms. 
More specifically, we choose the splitting in such a way that
the affine null parameter plays the role of time whose level surfaces are 3 dimensional spacelike hypersurfaces.

I also present the canonical formalism of the (2+2) bundle formulation of general
relativity, and  find that the constraint algebra is closed under 
the Poisson bracket. Thus, the constraints in the (2+2) formalism are 
the first-class constraints\cite{torre86,husain94,torre96}, 
just as the constraint algebra in the (3+1) formalism.
The (2+2) constraint algebra is the residual symmetry algebra of the spacetime metric 
adapted to the (2+2) splitting\cite{newman80}, and it is shown that it contains several interesting subalgebras such as the infinite dimensional Lie algebra of the diffeomorphisms of 2 dimensional vertical space,  an infinite dimensional
Virasoro algebra associated with the base manifold, and an analog of 
supertranslation.
As the constraint algebra is an infinite dimensional residual symmetry of
the spacetime at a finite distance, it may be regarded 
as {\it a} generalization of the asymptotic BMS symmetry algebra\cite{ashtekar81,kozameh85,brown86,ashtekar97,szabados03,barnich10}
or the Spi group\cite{chrusciel89,ashtekar92}.

In section \ref{s:metric}, the kinematics of the (2+2) splitting
of the spacetime metric is reviewed, and the Einstein's equations are presented. 
The affine null parameter associated with the out-going null congruence plays the role of time, and the splitting is chosen such that 
constant time slices define 3 dimensional spacelike hypersurfaces.
In section \ref{s:first}, Hamiltonian formalism is developed and Hamilton's 
equations of motion are derived from the first-order action principle,  which are found to be identical to the Einstein's equations.
In section \ref{s:constraint}, the complete set of the constraint algebra is presented,
and then discussion follows. 


\section{Spacetime metric in the (2+2) splitting and the Einstein's equations}
\label{s:metric}

In this section, I will introduce the kinematics of the (2+2) formalism\cite{yoon92,yoon93a,yoon99a,yoon04} of 4 dimensional spacetimes 
and present the Einstein's equations. The spacetime metric in the (2+2) splitting can be written as
\be \hspace{-1.8cm}
ds^2 = 2dudv - 2hdu^2 +\phi_{ab}
 \left( dy^a + A_{+}^{\ a}du +A_{-}^{\ a} dv \right) 
\left( dy^b + A_{+}^{\ b}du +A_{-}^{\ b} dv \right).                    \label{yoon}
\ee
Let $+,-$ stand for $u,v$, respectively, and denote 
\be
\partial_{+}={\partial \over \partial u}, \hspace{0.2cm}
\partial_{-}={\partial \over \partial v}, \hspace{0.2cm}
\partial_{a}={\partial  \over \partial y^{a}}
\hspace{0.2cm}  (a=2,3).
\ee
The hypersurface $v={\rm constant}$ is ruled by 
$\partial_{+} - A_{+}^{\ a}\partial_{a}$
whose norm is $-2h$, which can be either positive, zero, or negative.
In this paper, the sign is chosen 
\be
-2h>0                                              \label{near}
\ee
so that $v= {\rm constant}$ is a spacelike hypersurface. 
The hypersurface $u={\rm constant}$ is an out-going null hypersurface
ruled by the null vector field $\partial_{-} - A_{-}^{\ a}\partial_{a}$.
The intersection of two hypersurfaces $u,v={\rm constant}$ defines
a {\it compact} spacelike 2 surface $N_{2}$ labeled by $y^{a}$ 
with a positive-definite metric $\phi_{ab}$ on it (see FIG. 1). 
The metric $\phi_{ab}$ is decomposed into
the area element  $\tau$ and the conformal two metric
$\rho_{ab}$ normalized to have a unit determinant
\be
\phi_{ab}=\tau\rho_{ab}
\hspace{0.2cm}
( \det \rho_{ab}=1 ).                   \label{det}
\ee
In literatures of null hypersurface formalisms, the coordinates are often chosen such 
that $A_{-}^{\ a}=0$\cite{newman80}. In that case, $\partial_{-}$ becomes an out-going null vector field with $v$ the corresponding affine parameter.

\begin{figure}
\begin{center}
\vspace{-1.3in}
\includegraphics[width=5in]{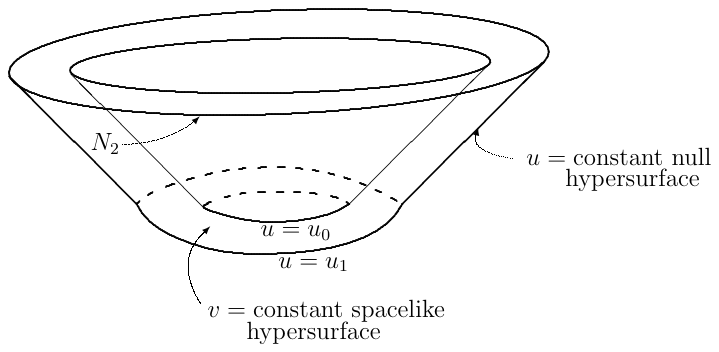}
\vspace{-10cm}
\caption{\label{figure1}Here the sign is chosen as $-2h>0$, so that
$v={\rm constant}$  is a spacelike hypersurface. The surface defined by 
$u={\rm constant}$ is an out-going null hypersurface, and 
$N_{2}$ is the cross section of two hypersurfaces $u,v={\rm constant}$, which we assume 
compact.}
\end{center}
\end{figure}

In order to compute the connections and the Ricci tensors, it is most convenient to introduce the horizontal vector fields
\begin{math}
\hat{\partial}_{\pm}
\end{math}
defined as
\be
\hat{\partial}_{\pm}:=\partial_{\pm} - A_{\pm}^{\ a}\partial_{a}. \label{pm}
\ee
The vector fields
\begin{math}
\hat{\partial}_{\pm}
\end{math}
are called horizontal since they are orthogonal to the vertical directions 
$\partial_{a}$. 
The basis $\hat{\partial}_{A}:=\{ \hat{\partial}_{\pm}, \partial_{a} \}$ 
is a non-coordinate basis, whose commutation relations are given by
\be
[ \ \hat{\partial}_{A}, \ \hat{\partial}_{B}] = f_{AB}^{ \ \ C}\,  \hat{\partial}_{C}.
\ee
In components, the only non-vanishing coefficients of $f_{AB}^{ \ \ C}$ are found to be
\bea
& & 
f_{+-}^{\ \ \ a}=-f_{-+}^{\ \ \ a}=-F_{+-}^{\ \ a },   \label{pub}\\
& &
f_{\pm a}^{\ \ \ b}=-f_{a \pm}^{\ \ \ b}= \partial_{a}A_{\pm}^{\ b},
\eea
where the field strength $F_{+-}^{\ \ a}$ is defined as
\bea
& &  \hspace{-1.2cm}
F_{+-}^{\ \ a}:=\partial_{+} A_{-} ^ { \ a}-\partial_{-} A_{+} ^ { \ a} 
- [A_{+}, \ A_{-}]^{a}   \no\\
& & \hspace{-0.5cm}
=\partial_{+} A_{-} ^ { \ a}-\partial_{-} A_{+} ^ { \ a} 
-A_{+}^{\ b}\partial_{b}  A_{-} ^ { \ a} 
+A_{-}^{\ b}\partial_{b}  A_{+} ^ { \ a}.  \label{throat}
\eea
The metric components in the non-coordinate basis are defined by the inner products 
of the basis vector fields, which are given by
\bea
& & 
\hat{\gamma}_{++}:=<\hat{\partial}_{+}, \ \hat{\partial}_{+}> = -2h, \label{candy}\\
& &
\hat{\gamma}_{+-}=\hat{\gamma}_{-+}
:=<\hat{\partial}_{+}, \ \hat{\partial}_{-}> = 1, \label{hope}\\
& &  
\hat{\gamma}_{--}:=<\hat{\partial}_{-}, \ \hat{\partial}_{-}> =0, \label{hall}\\
& &
\hat{\gamma}_{\pm a}=\hat{\gamma}_{a \pm}
:=<\hat{\partial}_{\pm}, \ \partial_{a}> =0,    \label{jungok}\\
& & 
\hat{\gamma}_{a b}:=<\partial_{a}, \ \partial_{b}> =\tau \rho_{ab}.
\eea
The Christoffel symbols and Ricci tensors are given by
\bea
& & \hspace{-2cm}
\hat{\Gamma}_{AB}^{\ \ \,C}={1\over 2}\hat{\gamma}^{CD} \big( 
\hat{\partial}_{A}\hat{\gamma}_{BD} +\hat{\partial}_{B}\hat{\gamma}_{AD}  
- \hat{\partial}_{D}\hat{\gamma}_{AB}  \big)
+{1\over 2}\hat{\gamma}^{CD} \big(  f_{ABD} - f_{BDA}- f_{ADB} \Big), \label{emma}\\
& &  \hspace{-2cm}
\hat{{\rm R}}_{AC}=\hat{\partial}_{A}\hat{\Gamma}_{BC}^{\ \ \ B} 
-\hat{\partial}_{B}\hat{\Gamma}_{AC}^{\ \ \ B} 
+\hat{\Gamma}_{AE}^{\ \ \ B} \hat{\Gamma}_{BC}^{\ \ \ E} 
-\hat{\Gamma}_{BE}^{\ \ \ B} \hat{\Gamma}_{AC}^{\ \ \ E} 
-f_{A B}^{ \ \ \ E}\hat{\Gamma}_{EC}^{\ \ \ B},
\eea
where $ f_{ABD}:=f_{AB}^{ \ \ \ C}\hat{\gamma}_{CD}$, and
$\hat{\gamma}^{AB}$ is the inverse of the metric $\hat{\gamma}_{AB}$
satisfying the relation
\be
\hat{\gamma}_{AB}\hat{\gamma}^{BC} = \delta_{A}^{\ C}.
\ee
The complete set of the Einstein's equations $\hat{\rm R}_{AB} =0$ 
is found as follows:
\begin{eqnarray}
 & & \hspace{-2cm} 
(a) \ \
2D_{-}^{2}\tau - \tau^{-1}(D_{-}\tau)^{2}
    + {\tau\over 2}\rho^{a b}\rho^{c d} (D_{-}\rho_{a c})
    (D_{-}\rho_{b d})   =0,                 \label{voice} 
\eea
\begin{eqnarray}
 & & \hspace{-2cm} 
(b) \ \ 
D_{-}(  \tau^{2} \rho_{a b}F_{+-}^{\ \ b})
+\partial_{a}(D_{-}\tau) - \tau^{-1}(D_{-}\tau)(\partial_{a} \tau)
+ {\tau\over 2}\rho^{b c}\rho^{d e}
    (D_{-}\rho_{b d})(\partial_{a}\rho_{c e}) \no\\
& & \hspace{-1.5cm} 
- \partial_{b} ( \tau \rho^{b c}D_{-}\rho_{a c} ) =0,\label{deeper} 
\eea
\begin{eqnarray}
 & & \hspace{-2cm} 
(c) \ \ 2 D_{-}^{2} h +2 \tau^{-1}(D_{-}h)(D_{-}\tau)
+ \tau^{-1} (D_{+}D_{-}\tau   
 + D_{-}D_{+}\tau)   \no\\
& &  \hspace{-1.5cm} 
-\tau^{-2}(D_{+}\tau)(D_{-}\tau)
+ {1\over 2}\rho^{a b}\rho^{c d}
  (D_{+}\rho_{a c})(D_{-}\rho_{b d})   
-\tau \rho_{a b}
    F_{+-}^{\ \ a}F_{+-}^{\ \ b} \no\\
  & & \hspace{-1.5cm} 
+ 2h \tau^{-1} \big\{
   D_{-}^{2} \tau  - {1\over 2}\tau^{-1}(D_{-}\tau)^{2}    
+{\tau\over 4}\rho^{a b}\rho^{c d}
   (D_{-}\rho_{a c})(D_{-}\rho_{b d})\big\}=0,  \label{repeat}
    \eea
\begin{eqnarray}
 & & \hspace{-2cm} 
(d) \ \
h\big\{\tau  D_{-}^{2} \rho_{ab}
- \tau \rho^{c d}(D_{-}\rho_{a c})(D_{-}\rho_{b d})
+(D_{-}\tau) (D_{-}\rho_{a b}) \big\}  \nonumber\\
& & \hspace{-1.5cm} 
+ {\tau\over 2}(
D_{+}D_{-}\rho_{a b} + D_{-}D_{+}\rho_{a b} ) 
-{\tau\over 2} \rho^{c d}\big\{
(D_{-}\rho_{a c})(D_{+}\rho_{b d})
+(D_{-}\rho_{b c})(D_{+}\rho_{a d}) \big\}  \nonumber\\
& &\hspace{-1.5cm} 
+ {1\over 2}\big\{ (D_{+}\tau) (D_{-}\rho_{a b})
+ (D_{-}\tau) (D_{+}\rho_{a b})  \big\}  
 +\tau (D_{-}h) (D_{-}\rho_{a b})  \no\\
 & &\hspace{-1.5cm} 
+{\tau^{2}\over 2}\rho_{a c}\rho_{b d}
 F_{+-}^{\ \ c}F_{+-}^{\ \ d}   
-{\tau^{2}\over 4}\rho_{a b}
 \rho_{c d}F_{+-}^{\ \ c}F_{+-}^{\ \ d}=0, \label{word}
\eea
\bea
& & \hspace{-2cm} 
(e) \ \
 2 D_{+}D_{-}\tau + \partial_{a} (\tau F_{+-}^{\ \ a})
+ 2 (D_{-}h)(D_{-}\tau) 
+ {\tau^{2}\over 2}\rho_{a b}
F_{+-}^{\ \ a}F_{+-}^{\ \ b}            \no\\
& &    \hspace{-1.5cm} 
-\tau R^{(2)}  
+ h \big\{ \tau^{-1} (D_{-}\tau)^{2} 
-{\tau\over 2}\rho^{a b}\rho^{c d}
 (D_{-}\rho_{a c})(D_{-}\rho_{b d})\big\}=0, \label{said}
 \eea
\begin{eqnarray}
 & & \hspace{-2cm} 
(f) \ \
 - D_{+}^{2}\tau +  {1\over 2} \tau^{-1}(D_{+}\tau)^{2}
 + (D_{-}h) (D_{+}\tau)  -(D_{+}h)(D_{-}\tau)     \no\\
& &  \hspace{-1.5cm} 
+2h (D_{-}h)(D_{-}\tau) 
-\tau F_{+-}^{\ \ a}\partial_{a}h  
-{\tau \over 4}\rho^{a b}\rho^{c d}
 (D_{+}\rho_{a c})(D_{+}\rho_{b d})  
+\partial_{a}( \rho^{a b}\partial_{b}h ) \no\\
& &  \hspace{-1.5cm} 
+h\big\{ \tau^{-1}  (D_{+}\tau)   (D_{-}\tau)
-{\tau\over 2} \rho^{a b}\rho^{c d} (D_{+}\rho_{a c}) (D_{-}\rho_{b d}) 
+{\tau^{2} \over 2}\rho_{a b}F_{+-}^{\ \ a}F_{+-}^{\ \ b}  \no\\
& & \hspace{-1.5cm} 
-\tau R^{(2)} \big\} 
+h^{2}\big\{ \tau^{-1} (D_{-}\tau)^{2} 
-{\tau\over 2}\rho^{a b}\rho^{c d} 
(D_{-}\rho_{a c}) (D_{-}\rho_{b d})\big\} =0,  \label{many}
\eea
\begin{eqnarray}
 & & \hspace{-2cm} 
(g) \ \
D_{+}(  \tau^{2} \rho_{a b}F_{+-}^{\ \ b})
- \partial_{a} ( D_{+}\tau )  
+2\tau^{-1} (h D_{-}\tau  + {1\over 2} D_{+}\tau)(\partial_{a} \tau )  \no\\
& & \hspace{-1.5cm} 
- \tau \rho^{b d}\rho^{c e} ( h  D_{-}\rho_{d e} 
+ {1\over 2}  D_{+}\rho_{d e}  ) (\partial_{a}\rho_{bc}) 
+\partial_{b}\big( 2h \tau 
\rho^{b c}D_{-}\rho_{c a}  + \tau \rho^{bc}D_{+}\rho_{ca} \big)  
\no\\
& & \hspace{-1.5cm} 
-2h \partial_{a}(D_{-}\tau) -2\tau \partial_{a}D_{-}h  =0.           \label{times}
\eea
Here $R^{(2)}$ is the scalar curvature of $N_{2}$, and the diff$N_{2}$-covariant 
derivatives $D_{\pm}f_{ab\cdots}$ of a tensor density
$f_{a b\cdots }$ with weight $w$ with respect to the diffeomorphisms 
on $N_{2}$ are defined by
\be
D_{\pm}f_{a b\cdots}:= \partial_{\pm}f_{a b\cdots}
-[A_{\pm}, \ f]_{a b\cdots},               \label{covdiff}
\ee
where the bracket $[A_{\pm}, \ f]_{a b\cdots }$ is the Lie
derivative of $f_{ab \cdots}$ along $A_{\pm}:=A_{\pm}^{\ a}\partial_{a}$
given by
\bea
& & \hspace{-1.5cm}
[A_{\pm}, \ f]_{a b\cdots}
=A_{\pm}^{\ c}\partial_{c}f_{ab\cdots}
+f_{cb\cdots}\partial_{a}A_{\pm}^{\ c}
+f_{ac\cdots}\partial_{b}A_{\pm}^{\ c}  
\cdots
+w (\partial_{c}A_{\pm}^{\ c})f_{ab\cdots}.           \label{liebra}
\eea
For instance, the diff$N_{2}$-covariant derivatives of
the area element $\tau$ and the conformal two metric
$\rho_{a b}$, which are a scalar density and  a tensor density with weight $1$ and $-1$
with respect to the diff$N_{2}$ transformations, respectively,
are given by
\bea 
& & \hspace{-1.5cm}
D_{\pm}\tau=\partial_{\pm}\tau
-A_{\pm}^{\ a}\partial_{a}\tau
-(\partial_{a} A_{\pm}^{\ a}) \tau,      \label{walleye}\\
& &   \hspace{-1.5cm}
D_{\pm}\rho_{a b}=\partial_{\pm} \rho_{a b}
-A_{\pm}^{\ c}\partial_{c} \rho_{a b}
-\rho_{c b}\partial_{a}A_{\pm}^{\ c}
-\rho_{a c}\partial_{b}A_{\pm}^{\ c} 
+ (\partial_{c}A_{\pm}^{\ c})\rho_{a b}.               \label{bass}
\eea
On the other hand, the function $h$ is a scalar field under the diff$N_{2}$ transformations, 
so that the diff$N_{2}$-covariant derivative of $h$ is given by
\be
D_{\pm}h= \partial_{\pm}h - A_{\pm}^{\ a}\partial_{a}h.   \label{cut}
\ee
Notice that the equations (\ref{said}), (\ref{many}), and (\ref{times}) are
first-order equations in $D_{-}$ derivatives, whereas the remaining equations are 
second-order in  $D_{-}$ derivatives. 
Thus, it is natural to regard $D_{-}$ as the {\it time} derivative 
in this (2+2) formalism. Then the  equations (\ref{said}), (\ref{many}), and (\ref{times}) 
are the constraint equations defined on a $v={\rm constant}$ spacelike hypersurface,
and the equations (\ref{voice}), (\ref{deeper}), (\ref{repeat}), and (\ref{word})  are the evolution equations with respect to the $D_{-}$ derivative.


\section{First-order form of the Einstein's equations}
\label{s:first}

Let us {\it define} the canonical momenta 
$\pi_{I}=(\pi_{\tau}, \pi_{h},\pi_{a}, \pi^{ a b})$ conjugate to the configuration variables
$q^{I}=(\tau, h, A_{+}^{\ a}, \rho_{ab})$ as the $D_{-}$ derivatives
of the configuration variables as follows,
\bea
& & 
\pi_{h}:=2D_{-}\tau,   \label{broker}\\
& &
\pi_{a}:=-\tau^{2} \rho_{a b}F_{+-}^{\ \ b}, \label{heart}\\
& &
\pi_{\tau} :=2h \tau^{-1}D_{-}\tau +2D_{-}h  + \tau^{-1}D_{+}\tau , \label{jack}\\
& &
\pi^{a b}:=-h \tau \rho^{a c}\rho^{b d}D_{-}\rho_{c d}
-{1\over 2}\tau \rho^{a c}\rho^{b d}D_{+}\rho_{c d}.      \label{after}
\eea
Then the constraint equations (\ref{said}), (\ref{many}), and (\ref{times}) 
in the $(q^{I}, \pi_{I})$ variables become
\begin{eqnarray}
& & \hspace{-2cm}  
(1) \
C_{-}:=  {1\over 2}\pi_{h}\pi_{\tau}
-{h\over 4\tau}\pi_{h}^{2}
-{1\over 2\tau}\pi_{h}D_{+}\tau
+{1\over 2\tau^{2}}\rho^{a b}\pi_{a}\pi_{b} 
- \partial_{a}(\tau^{-1}\rho^{a b} \pi_{b})  -\tau R^{(2)} \nonumber\\
& &   \hspace{-1.5cm} 
-{\tau\over 8h}\rho^{a b} \rho^{c d}
(D_{+}\rho_{a c}) (D_{+}\rho_{b d})
-{1\over 2h \tau}
\rho_{a b}\rho_{c d}\pi^{a c}\pi^{b d}  
-{1\over 2h}\pi^{a c}D_{+}\rho_{a c}
+ D_{+}\pi_{h} \nonumber\\
& &  \hspace{-1.5cm} 
 =0,                      \label{step}
\end{eqnarray}
\begin{eqnarray}
& & \hspace{-2cm} 
(2) \ 
C_{+}:=\pi_{\tau}D_{+}\tau + \pi_{h}D_{+}h + \pi^{a b}D_{+}\rho_{a b} 
 - 2D_{+}( h \pi_{h}  +  D_{+} \tau )  \nonumber\\
& &   \hspace{-1.5cm} 
+2\partial_{a}(h \tau^{-1}\rho^{a b}\pi_{b}
 +\rho^{a b}\partial_{b}h ) =0,                     \label{horse}
\end{eqnarray}
\begin{eqnarray}
& & \hspace{-2cm} 
(3) \
C_{a}:=\pi_{\tau}\partial_{a}\tau +\pi_{h}\partial_{a}h
+\pi^{b c}\partial_{a}\rho_{b c}
-2\partial_{b}( \rho_{ac}\pi^{bc})   -D_{+}\pi_{a} - \partial_{a}(\tau \pi_{\tau}) =0. \label{sharp}
\end{eqnarray}
The corresponding first-order action principle is 
\be
 S=\int \!\! dv du d^{2} y \{ \pi_{\tau}\p_{-}{\tau} + \pi_{h}\p_{-}{h}
+ \pi_{a}\p_{-}{A}_{+}^{\ a}   
+ \pi^{ab}\p_{-}{\rho}_{ab}  -K \},          \label{act}
\ee
where $K$ is  the sum of the constraints given by
\be
K=  ``1" \cdot C_{-} +``0" \cdot C_{+} + A_{-}^{\ a}C_{a},  \label{kei}
\ee
and  $``1"$, $``0"$, $A_{-}^{\ a}$ are the Lagrange multipliers associated with
the constraints $C_{-}=0$, $C_{+}=0$, and $C_{a}=0$, respectively.
The diff$N_{2}$-covariant derivatives of the conjugate momenta $\pi_{\tau}$, 
$\pi_{h}$, $\pi_{a}$, and $\pi^{ab}$, which are tensor densities of weight
$0, 1, 1, 2$, respectively, are given by 
\bea 
& &  \hspace{-1cm}
D_{\pm} \pi_{\tau}=\partial_{\pm} \pi_{\tau}  
-  A_{\pm}^{\ a}\partial_{a}\pi_{\tau},                         \label{close}\\
& & \hspace{-1cm}
 D_{\pm} \pi_{h}=\partial_{\pm} \pi_{h}  
-  A_{\pm}^{\ c}\partial_{c}\pi_{h} -(\partial_{c} A_{\pm}^{\ c}) \pi_{h}, \label{hill}\\
& & \hspace{-1cm}
D_{\pm} \pi_{a}=\partial_{\pm} \pi_{a}  
-  A_{\pm}^{\ c}\partial_{c}\pi_{a}  -\pi_{c}\partial_{a}A_{\pm}^{\ c}
- (\partial_{c} A_{\pm}^{\ c}) \pi_{a},       \label{bank}\\
& & \hspace{-1cm}
D_{\pm} \pi^{ab}=\partial_{\pm}\pi^{ab}
-  A_{\pm}^{\ c}\partial_{c}\pi^{ab}   +\pi^{cb}\partial_{c}A_{\pm}^{\ a}
+ \pi^{ac}\partial_{c}A_{\pm}^{\ b}  
-2 (\partial_{c} A_{\pm}^{\ c}) \pi^{ab}.  \label{deep}
\eea
Hamilton's equations of motion are obtained by varying $K$ with respect to
$\rho_{ab}$ and $\pi^{ab}$ subject to the condition 
\be
\det \rho_{ab} =1.                      \label{deter}
\ee
This can be done by adding another Lagrange multiplier $\lambda$ enforcing the
kinematic constraint (\ref{deter}) to the Hamiltonian $K$. Define $\tilde{K}$ as
\be
\tilde{K}:=K +  \lambda (\det \rho_{ab}-1),
\ee
then Hamilton's equations of motion are given by
\bea 
& &  \hspace{-0.8cm}
\p_{-}q^{I}=\int_{\Sigma_{3}}\!\! \!\! du d^{2} y 
  {\delta  \tilde{K} \over \delta \pi_{I}},     \label{push}\\
& & \hspace{-0.8cm}
\p_{-}\pi_{I}=-\int_{\Sigma_{3}}\!\! \!\! du d^{2} y 
{\delta  \tilde{K}\over \delta q^{I}}.  \label{chin}
\eea
Notice that, in the above action principle, the functional derivatives with respect to $\pi_{I}$ are well-defined if the boundary condition on the Lagrange multiplier 
\be
A_{-}^{ \ a}=0
\ee
is assumed on the spatial boundary $\partial \Sigma_{3}$
(assuming that $N_{2}$ is compact), and that the functional derivatives
with respect to $q^{I}$ are well-defined if the boundary conditions 
\be
\delta \tau =0, \  \
\delta \rho_{ab}=0
\ee
are assumed on $\partial \Sigma_{3}$.

The first half of Hamilton's equations of motion (\ref{push})
reproduces $D_{-}$ derivatives of $q^{I}$,
\bea
& & 
(4) \ 
D_{-}\tau
=  {1\over 2}\pi_{h},                    \label{real}\\
& & 
(5) \ 
F_{+-}^{\ \ a}=-\tau^{-2} \rho^{a b}\pi_{b},    \label{ima}\\
& & 
(6) \ 
D_{-}h =  {1\over 2}\pi_{\tau}
 -{1\over 2\tau}D_{+}\tau
 -{1\over 2\tau}h\pi_{h},                  \label{leaf}\\
& & 
(7) \ 
D_{-}\rho_{a b}=-{1\over \tau h}
\rho_{a c}\rho_{b d}\pi^{c d}
-{1\over 2h}D_{+}\rho_{a b},             \label{little}
\eea
which are {\it precisely} the equations (\ref{broker}), (\ref{heart}), (\ref{jack}), and (\ref{after}).
By taking the derivative of the uni-modular condition (\ref{deter}),  we find that
\begin{equation}
\rho^{bc}\partial_{\pm}\rho_{bc}
=\rho^{bc}\partial_{a}\rho_{bc}=\rho^{bc}D_{\pm}\rho_{bc}=0,  \label{module}
\end{equation}
and from this equation and the equation (\ref{little}), $\pi^{a b}$ is traceless,
\begin{equation}
\rho_{ab}\pi^{a b}=0.          \label{side}
\end{equation}
The other half of Hamilton's equations of motion (\ref{chin})
defines $D_{-}$ derivatives of the momenta $\pi_{I}$,
\bea  
& & \hspace{-2cm}
(8) \  D_{-}\pi_{h}
 =   {1\over 4\tau}\pi_{h}^2
-{1\over 2 \tau h^2}\rho_{a b} \rho_{c d}
\pi^{a c}\pi^{b d}
-{1\over 2 h^2}\pi^{a b}D_{+}\rho_{a b}  \no\\
& & \hspace{-0.6cm}
-{\tau\over 8h^2}  \rho^{a b}\rho^{c d}
(D_{+}\rho_{a c})(D_{+}\rho_{b d}),               \label{pile}\\
& &  \hspace{-2cm}
(9) \  
D_{-}\pi_{a} =  -{1\over 2\tau}\pi_{h}\partial_{a}\tau
+{1\over 2}\partial_{a}\pi_{h} 
- {1\over 2h} \pi^{b c} \partial_{a}\rho_{bc}   
-{\tau\over 4h}
\rho^{b d}\rho^{c e}(D_{+}\rho_{d e} ) (\partial_{a}\rho_{bc})     \no\\
& &   \hspace{-0.6cm}
+\partial_{b}  ({1\over h}\pi^{b c} \rho_{c a})   
+\partial_{b} ( {\tau\over 2h}
\rho^{b c}D_{+}\rho_{c a}),           \label{hot}\\
& &  \hspace{-2cm}
(10) \
D_{-}\pi_{\tau} = -{1\over 4\tau^{2} }h\pi_{h}^{2}
-{1\over 2\tau}D_{+}\pi_{h}
+{1\over \tau^{3}}  \rho^{a b}\pi_{a} \pi_{b}  
-{1\over 2\tau^{2} h}\rho_{a b} \rho_{c d}
\pi^{a c}\pi^{b d}       \no\\
& &  \hspace{-0.6cm}
+{1\over 8h} \rho^{a b}\rho^{c d}
(D_{+}\rho_{a c})(D_{+}\rho_{b d}),         \label{fire}\\
& & \hspace{-2cm}
(11) \  
D_{-}\pi^{a b} =   {1\over 2\tau^{2}}
\rho^{a c}\rho^{b d}\pi_{c} \pi_{d}
+{1\over \tau h}\rho_{c d}
\pi^{a c}\pi^{b d}            
-{\tau\over 4h}\rho^{a c}\rho^{b d}\rho^{e f}
(D_{+}\rho_{c e})(D_{+}\rho_{d f})  \nonumber\\
& & \hspace{-0.6cm}
-D_{+}\big\{ {1\over 2h}\pi^{a b}
+{\tau \over 4h}\rho^{a c}\rho^{b d}
(D_{+}\rho_{c d} ) \big\} - \lambda \rho^{ab}. \label{gray}
\eea
By taking trace of the equation (\ref{gray}) with $\rho^{ab}$, we find that
the Lagrange multiplier $\lambda$ is determined as
\be
\lambda = {1\over 4\tau^{2}}\rho^{cd} \pi_{c}\pi_{d}.  \label{mul}
\ee
With this value of
$\lambda$, the equation (\ref{gray}) becomes
\bea
& &  \hspace{-2cm}
(11') \  
D_{-}\pi^{a b}  =   {1\over 2\tau^{2}}
\rho^{a c}\rho^{b d}\pi_{c} \pi_{d}
-{1\over 4\tau^{2}}
\rho^{a b}\rho^{c d}\pi_{c} \pi_{d} 
+{1\over \tau h}\rho_{c d} 
\pi^{a c}\pi^{b d}               \nonumber\\
& & \hspace{-1.2cm}
-{\tau\over 4h}\rho^{a c}\rho^{b d}\rho^{e f}
(D_{+}\rho_{c e})(D_{+}\rho_{d f}) 
-D_{+}\big\{ {1\over 2h}\pi^{a b}
+{\tau \over 4h}\rho^{a c}\rho^{b d}
(D_{+}\rho_{c d} ) \big\}.    \label{state}
\eea
After some algebra, it is easy to show that, from the equations  (\ref{real}), $(\ref{ima})$, (\ref{leaf}), (\ref{little}),  $(\ref{pile})$, (\ref{hot}), (\ref{fire}), and (\ref{state}), one can reproduce the evolution equations (\ref{voice}), (\ref{deeper}), (\ref{repeat}), and (\ref{word}). Thus, from the first-order form of the action principle (\ref{act}), 
we obtain the four constraint equations  (\ref{step}), (\ref{horse}), (\ref{sharp}) 
associated with the Lagrange multipliers $``1"$, $``0"$, $A_{-}^{\ a}$, respectively,
and Hamilton's equations of motion (\ref{push}) and (\ref{chin}) by varying with respect to 
$(q^{I}, \pi_{I})$. These equations, when put together, 
are identical to the Einstein's equations 
\be
\hat{{\rm R}}_{AB}=0                    \label{rich}
\ee
for the spacetime metric (\ref{yoon}).

\section{Constraint algebra in the (2+2) formalism}
\label{s:constraint}

In this section, I will present the algebra satisfied by the constraints 
$C_{\pm}=0$ and $C_{a}=0$ on a given spacelike 
hypersurface ${\Sigma_{3}}$ defined by $v=$ constant. 
Let us introduce the classical Poisson bracket
\be
\{  F, \ G \}_{\rm P} :=\int_{\Sigma_{3}}\!\! \!\! du d^{2} y \big(
{\delta  F\over \delta q^{I}} {\delta  G\over \delta \pi_{I}}
-{\delta  F\over \delta \pi_{I}} {\delta  G\over \delta q^{I} }   \big)
\ee
for functions $F$ and $G$ defined on the phase space $( q^{I}, \pi_{I})$.
Let $\xi^{a}$ and $f$ be suitable test functions of $(u,y^{a})$ 
with a compact support, and define functionals 
$C{(\xi)}$ and $C_{\pm}(f)$ as
\bea
& & C{(\xi)}:=\int_{\Sigma_{3}}\!\!  \!\! du d^{2} y \, \xi^{a} C_{a},  \label{love}\\
& & C_{\pm}(f):=\int_{\Sigma_{3}}\!\!  \!\! du d^{2} y \, f \, C_{\pm}. \label{what}
\eea
After lengthy but straightforward calculations, it is found that 
these functionals are closed under the Poisson bracket and satisfy the following algebra,
\bea
& & \hspace{-1.5cm}
\{ C(\xi), \ C(\eta) \}_{\rm P}=C( [ \xi, \eta ] ),    \label{ben}\\
& &  \hspace{-1.5cm}
\{ C_{+}(f), \ C_{+}(g) \}_{\rm P} 
=C_{+}(f D_{\! +}g -g D_{\! +}f) \no\\
& & \hspace{1cm}
+\int_{\Sigma_{3}}\!\!  \!\! du d^{2} y \, 
2h \tau^{-1}\rho^{ab} (f \partial_{b}g -g\partial_{b}f ) C_{a}, \label{mike}\\
& &  \hspace{-1.5cm}
\{ C(\xi), \ C_{+}(f) \}_{\rm P}= C_{+}(\mathcal{L}_\xi  f),   \label{aga}\\
& &  \hspace{-1.5cm}
\{ C_{-}(f), \ C_{-}(g) \}_{\rm P} =0 \hspace{0.2cm} {\rm (strictly)},  \label{yes}\\
& &  \hspace{-1.5cm}
\{  C_{+}(f), \ C_{-}(g) \}_{\rm P}= C_{-}( D_{+}(fg))  \nonumber\\
& &  \hspace{1cm}
+\int_{\Sigma_{3}}\!\!  \!\! du d^{2} y \,  
\tau^{-1}\rho^{ab}\big[  ( f\partial_{b}g -g\partial_{b} f) 
+ \tau^{-1}\pi_{b}\, (f  g ) 
\big]
C_{a},  \label{huge}\\
& &  \hspace{-1.5cm}
\{  C(\xi), \ C_{-}(f) \}_{\rm P}= C_{-} (\mathcal{L}_\xi f ),  \label{shake}
\eea
where
\bea
& & 
[ \xi, \eta ]^{a} =\xi^{b}\partial_{b}\eta^{a}-\eta^{b}\partial_{b}\xi^{a}, \label{why}\\
& & 
\mathcal{L}_\xi f =\xi^{a}\partial_{a}f,           \label{haag}\\
& & 
D_{+}f=\partial_{+}f - A_{+}^{\ a}\partial_{a} f.           \label{expect}
\eea 
Therefore, the constraints in the (2+2) formalism 
are first-class constraints, just as the Hamiltonian and momentum 
constraints in the standard ADM formalism. Because the constraints (\ref{step}),  
(\ref{horse}), and (\ref{sharp}) are written in the specific coordinates in which the 
metric assumes the form (\ref{yoon}),  they are responsible for
the residual symmetries of the metric (\ref{yoon}).

There are several interesting subalgebras of the above algebra.  First, notice that 
the subalgebra (\ref{ben}) is the 
infinite dimensional Lie algebra of the diffeomorphisms of 2 surface $N_{2}$,
\be
y^{a} \longrightarrow y'^{a}=y'^{a}(u, y^{b}).
\ee
This corresponds to the fact that any diffeomorphic coordinate transformations 
on $N_{2}$ with an arbitrary $u$-dependence preserves the symmetry of the metric\cite{newman80}. 

The algebra (\ref{mike}) contains
the infinite dimensional Virasoro algebra as a subalgebra, as can be seen as follows. 
Assume that the test functions $f$ and $g$ are functions of $u$ only, 
so that 
\be
\partial_{a}f= \partial_{a}g= 0.
\ee
If we expand them in a Fourier series
\be
f(u)=\sum_{m}f_{m} e^{i m u}, \hspace{0.2cm}
g(u)=\sum_{n}g_{n} e^{i n u},
\ee
then, we have
\bea
& & 
C_{+}(f)=\sum_{m}f_{m}L_{m},  \label{cart}\\
& &
C_{+}(g)=\sum_{n}g_{n}L_{n},   \label{carry}\\
& & 
C_{+}(fD_{\! +}g - gD_{\! +}f) 
= i \sum_{m,n}(n-m)f_{m}g_{n} L_{m+n},
\eea
where 
\be
L_{m}:=\int_{\Sigma_{3}}\!\!   \!\! du d^{2} y \, e^{i m u} C_{+}.
\ee
It follows that the algebra (\ref{mike}) reduces to the Virasoro algebra\cite{husain94,torre96,barnich10}
\be
\{ L_{m}, \ L_{n} \}_{\rm P} =  i (n-m) L_{m+n}.      \label{virasoro}
\ee
Thus, the constraint $C_{+}=0$ generates an infinite dimensional algebra that contains
the Virasoro algebra as a subalgebra, which is associated with the following
reparametrization  of $u$, or equivalently, a deformation of $u={\rm constant}$ subspace on $\Sigma_{3}$\cite{newman80},
\be
u \longrightarrow u'=f (u). 
\ee
In order to preserve the metric, however, the reparametrization of $u$ must be accompanied by the corresponding scale transformation of $v$,
\be
v \longrightarrow v' =   (d f / d u)^{-1}  v. 
\ee
Modulo the diffeomorphisms of $N_{2}$ generated by the $C_{a}=0$ constraint,  
the algebra (\ref{mike}) becomes
\be
\{ C_{+}(f), \ C_{+}(g) \}_{\rm P} 
\approx C_{+}(f D_{\! +}g - g D_{\! +}f  ).  \label{michael}
\ee
In general, the algebra (\ref{michael}) should be viewed as 
a  ``gauged"  or ``vertical" deformation of the Virasoro algebra, in the sense that 
the diff$N_{2}$-covariant derivative $D_{+}$ or the ``vertical lift" of $\partial_{+}$ is placed instead of $\partial_{+}$ in the $C_{+}=0$ constraint. 

There is another interesting subalgebra (\ref{yes}), which is Abelian,
as it commutes for any functions $f(u,y^{a})$ and $g(u,y^{a})$. The residual symmetry associated with this Abelian subalgebra  is the {\it shift of the origin} of the out-going null congruences\cite{newman80}
\be
v \longrightarrow v'= v + \alpha (u,y^{a})                              \label{fist}
\ee
by any function $\alpha(u,y^{a})$ at each point on $\Sigma_{3}$.
Because the affine parameter $v$ of the out-going null congruence plays the role of 
time in this (2+2) formalism, the transformation (\ref{fist}) generates 
the symmetry of an arbitrary $y^{a}$-dependent ``time" translation
at a finite distance.  Thus, the transformation (\ref{fist}) may be regarded as
an analog or a generalization of the {\it supertranslation} 
in the asymptotic BMS or Spi symmetry group to a finite distance.

\section{Summary}
\label{s:summary}

In this paper,  I presented the  first-order form of the Einstein's equations in
the (2+2) formalism as a set of Hamilton's equations of motion 
and the constraint equations. The initial data are given on a spacelike hypersurface $\Sigma_{3}$ defined by $v={\rm constant}$, and the evolution is defined 
with respect to the out-going null vector field. Assuming the Poisson bracket 
\be
\{ q^{I}(x), \ \pi_{J}(x') \}_{\rm P} =\delta^{I}_{J}\delta^{(3)} (x, x')
\hspace{1cm}
(x,x' \in \Sigma_{3}),
\ee
I have shown that the constraint algebra is closed
under the Poisson bracket. The symmetries generated by the constraints are 
realizations of the residual symmetries of the spacetime metric (\ref{yoon}), which 
contains as subalgebras the infinite dimensional Lie algebra of the diffeomorphisms of 
the vertical 2 surface,  an infinite dimensional Virasoro algebra, and 
an Abelian subalgebra. As the infinite dimensional symmetry generated by 
the constraints is  a residual symmetry preserving the local structure 
of the spacetime metric (\ref{yoon}), it may be viewed as a finite-distance generalization of 
the BMS or the Spi symmetry group preserving asymptotic structures at infinities.

It must be mentioned that the (2+2) canonical formalism is local in nature, because
the evolution is defined with respect to the affine parameter of the out-going null vector field at each point of the 3 dimensional spacelike hypersurface. In general, caustics 
could appear in a finite affine time, and then the coordinates become singular at the caustic surfaces. This seems unavoidable, unless one replaces the affine parameter 
that plays the role of time with a suitable function in the phase space 
by a procedure such as Hamiltonian reduction\cite{ADM60,kuchar71}. Hamiltonian
reduction of the Einstein's theory using the (2+2) canonical formalism 
is under progress\cite{yoon2014}. 
On the other hand, the (2+2) formalism exhibits 
a finer canonical structure, as is discussed above,
which is hard to observe in the standard ADM formalism. Moreover, the Hamiltonian constraint is more ``tamed", because the 2 dimensional scalar curvature $R^{(2)}$,
which becomes the Euler number after integration over a 2 surface,
appears in the Hamiltonian constraint instead of the 3 dimensional scalar curvature $R^{(3)}$.
Thus, the (2+2) canonical formalism provides a complementary 
way of looking at the constraints of the Einstein's theory.

A related question is to find the relationship of the phase space of the (2+2) canonical formalism and the standard ADM phase space. Because the out-going null vector field that
defines the evolution in the (2+2) formalism may be viewed as the singular limit of timelike vector fields, results obtained in the (2+2) canonical formalism may also be
obtained from the corresponding results in the (3+1) formalism 
in an appropriate limit. Conversely, by introducing a unit-normal timelike 
vector field as a combination of in- and out-going null vector fields, it should be possible to express the conjugate momenta in the (3+1) formalism in terms of the phase space variables in the (2+2) formalism. 
This will define a natural symplectic map between the two canonical formalisms.  
To show the equivalence of the two canonical formalisms by establishing the existence of 
the symplectic transformation of this kind, and then reconstructing 
the standard Hamiltonian and momentum constraints out of the (2+2) 
constraints presented in this paper is an interesting problem left for future works\cite{rendall90,choquet09,brill87}.


\section*{Acknowledgments}
The author thanks the referees for valuable and constructive comments, especially 
for pointing out the possible existence of a symplectic map between the phase spaces
of the two formalisms.
This paper was written as part of Konkuk University's research support program
for its faculty (2011-A019-0035).

\end{document}